\documentclass[lettersize,journal]{IEEEtran}
\usepackage{amsmath,amsfonts}
\usepackage{algorithmic}
\usepackage{algorithm}
\usepackage{array}
\usepackage[caption=false,font=normalsize,labelfont=sf,textfont=sf]{subfig}
\usepackage{textcomp}
\usepackage{stfloats}
\usepackage{url}
\usepackage{verbatim}
\usepackage{graphicx}
\usepackage{cite}
\usepackage{hyperref}       
\usepackage{caption}
\usepackage[x11names]{xcolor}

\begin{document}

\title{Beyond Per-Request QoS: Coordinating Industrial Workflows with B5G/6G Network Capabilities}

\author{Qize~Guo,
Bjoern~Riemer,
Tarik~Taleb,
 Yan~Chen, 
 Hao~Yu,
 and Hemant~Zope

\thanks{Qize~Guo, Yan~Chen, and Tarik~Taleb are with the Faculty of Electrical Engineering and Information Technology, Ruhr University Bochum, Bochum, 44801, Germany. e-mail: {Qize.Guo@rub.de, yanchen@ieee.org, tarik.taleb@rub.de}.}
\thanks{Bjoern~Riemer and Hemant~Zope are with the Fraunhofer Institute for Open Communication Systems (FOKUS), Berlin, 10589, Germany. e-mail: bjoern.riemer@fokus.fraunhofer.de, hemant.zope@fokus.fraunhofer.de.}
\thanks{Hao~Yu is with the Hangzhou International Innovation Institute, Beihang University, Hangzhou, 311115, China. Email: haoyu1@buaa.edu.cn. (\textit{Corresponding author: Yan Chen})}}

\markboth{}%
{Guo \MakeLowercase{\textit{et al.}}: Beyond Per-Request QoS: Coordinating Industrial Workflows with B5G/6G Network Capabilities}

\bstctlcite{IEEEexample:BSTcontrol}
\bibliographystyle{IEEEtran}

\maketitle

\begin{abstract}
Beyond-5G (B5G) and 6G networks are expected to enable more complex industrial services, which often operate according to multi-phase workflows with phase-specific communication requirements. However, current interaction between applications and networks remains predominantly request-driven: Quality of Service (QoS) is requested at each workflow phase transition and evaluated independently, without explicit consideration of upcoming demand or the network's near-term capability. This mismatch limits the ability of both sides to plan ahead, often resulting in foreseeable incompatibilities, even service disruptions.
This article presents a capability-aware coordination framework for workflow-based industrial services. Within a bounded planning window, the network exposes the QoS profiles it can sustainably support, while the industrial side maps upcoming workflow phases to these disclosed capabilities and submits the resulting demand trajectory for joint assessment. The framework also supports coordinated updates when network conditions change during execution.
An industrial video inspection case study on an operational 5G/B5G testbed, complemented by large-scale simulation, illustrates that such coordination can improve service continuity, reduce disruptive rejections, and increase workflow completion under heavy load. The results suggest that future industrial networking should move beyond reactive per-request QoS handling toward forward-looking, capability-aware, workflow-level coordination.
\end{abstract}


\section{Introduction}
\IEEEPARstart{I}{ndustrial} operations across manufacturing, logistics, energy, and other industrial verticals are increasingly organized as structured workflows comprising multiple phases. Within these workflows, network services with assured Quality of Service (QoS) are increasingly indispensable for the reliable execution of phases such as inspection, control, real-time collaboration, and closed-loop operations~\cite{intro:5gppp2023InnovationTrends,intro:magon6gindustry}. However, as these workflows progress, the QoS requirements often vary significantly between phases~\cite{intro:camdynamicQoS,charpentierNextGenConnectivityDynamic2024}. For example, a robotic inspection workflow may involve lightweight telemetry during routine patrol, high-throughput video during active inspection, and moderate bandwidth during reporting. In such workflows, the overall success of the process hinges on the coordinated completion of all phases, rather than on the isolated completion of individual phases.

Beyond-5G (B5G) and emerging 6G networks provide strong support for differentiated QoS in network slicing through mechanisms such as policy control and service exposure~\cite{3gpp29522}. These mechanisms are well suited for individual QoS decisions at specific workflow phases. However, available network capability usually evolves over time due to fluctuating traffic load, varying radio conditions, and dynamic resource sharing among concurrent services. Meanwhile, industrial QoS demand also evolves as the workflow progresses from one phase to the next. Under current practice, QoS interaction remains predominantly request-driven, whereby applications submit new requests at phase transitions, and the network evaluates them against the instantaneous network capability. This request-driven paradigm can determine whether the requested QoS is supportable at the current phase, but cannot guarantee the feasibility of QoS required by subsequent workflow phases. To mitigate the impact of this limitation, several concepts and mechanisms have been developed. For example, the 3rd Generation Partnership Project (3GPP) Alternative QoS Profile~\cite{3gpp23503} supports fallback to pre-configured profiles when the current one becomes unsustainable. While effective for reactive degradation handling, it neither discloses which profiles are expected to remain supportable in the near term nor accepts QoS requests expressed over a sequence of upcoming workflow phases. Therefore, these mechanisms do not resolve the limitation. To address this issue, QoS interaction needs to evolve toward proactive coordination between near-term network capability and evolving workflow demand. The network should expose bounded near-term capability, while the industrial application should express the evolution of its QoS demands across upcoming phases, including acceptable degradation options.

This article introduces a capability-aware coordination framework for forward-looking QoS interaction in workflow-based industrial services, targeting current B5G systems and their evolution toward 6G networks. The framework is structured into three key constructs. A \emph{planning window} defines the finite future interval over which coordination is carried out. Within this window, the network exposes a \emph{capability envelope}, namely the set of QoS profiles it expects to sustain together with their validity bounds. Based on this exposed capability view, the industrial application side constructs a \emph{demand trajectory} that maps upcoming workflow phases to feasible profiles and specifies acceptable fallback options when the preferred profile cannot be sustained. The framework therefore enables workflow-level feasibility assessment and coordinated adaptation across multiple upcoming phases, rather than treating each QoS request as an isolated decision. 

The main contributions of this article are as follows:
\begin{itemize}
\item We identify the key limitation of current request-driven QoS interaction for industrial workflows: it provides no bounded forward-looking coordination scope within which evolving workflow demand and near-term network capability can be considered jointly.

\item We propose a capability-aware coordination framework built on three constructs, namely a planning window, a capability envelope, and a demand trajectory, to support workflow-level feasibility assessment and structured adaptation across upcoming workflow phases.

\item Through a case study on B5G-based industrial video inspection and large-scale simulations, we demonstrate that capability-aware coordination improves service continuity and workflow completion compared with conventional request-driven operation, especially under heavy traffic.
\end{itemize}


\section{Background and Related Work}
\label{sec:background}

We focus on a specific coordination problem between the vertical industry and the mobile network: how to align evolving workflow QoS demand with near-term network capability across multiple upcoming phases. This problem arises when workflow phases impose different QoS requirements and network capability varies over time. Existing work addresses parts of this problem, but lacks workflow-level, forward-looking coordination between the evolution of
industrial QoS demand and projected network capability.

One relevant line of work is resource reservation and service renegotiation. Protocols such as Resource Reservation Protocol (RSVP)~\cite{rsvp_rfc2205} establish per-flow guarantees through signaling across forwarding elements. Service Level Agreement (SLA) lifecycle and renegotiation approaches~\cite{sla_lifecycle} similarly support bilateral adjustment of service parameters. These mechanisms are important for reserving or modifying QoS profiles, but they are not designed to coordinate upcoming workflow phases against a bounded view of near-term network capability.

A second relevant line of work is intent-driven and agentic networking~\cite{ieeeintentsurvey2026,masoudmagazine,xiaoAgenticAINetworking2025b}. These approaches improve how objectives of applications are expressed, interpreted, and translated into management actions. Their main contribution lies in abstraction and automation. By contrast, the issue addressed in this article is how to coordinate projected network capability with the QoS evolution of a workflow during execution. In this sense, the proposed framework is complementary: it provides a structured coordination layer within which higher-level intent or agentic functions may operate.

The proposed framework is aligned with recent 3GPP developments. The Network Exposure Function (NEF) and Policy Control Function (PCF) provide the architectural foundation for application-network interaction~\cite{3gpp29522}. Predictive QoS notification, enabled by the Network Data Analytics Function (NWDAF)~\cite{3gpp23288}, allows the network to anticipate QoS changes for an active session and alert the application in advance. However, this mechanism remains unidirectional and session-scoped: it informs the application of possible degradation, but neither accepts upcoming demand information from the application nor supports workflow-level assessment across multiple phases. Alternative QoS Profiles~\cite{3gpp23503} support reactive fallback when the current profile becomes unsustainable, but do not disclose which profiles are expected to remain supportable in the near term. At the API level, Quality on Demand (QoD) enables applications to request QoS treatment for individual sessions~\cite{camara_qod}. These mechanisms are important enablers, but they do not provide a bounded view of near-term supportability across upcoming workflow phases.
 
The target scope of the proposed framework is consistent with many industrial deployments. We focus on workflows with known phase structures and phase-specific QoS requirements, such as scheduled inspection, periodic patrol, and batch processing. We also assume operation in fixed or well-controlled coverage, as is common in factories and campus networks, together with near-term network capability that can be projected from current state and committed demand. These assumptions make bounded capability disclosure meaningful while remaining realistic for the intended application domain. The framework is less suitable for highly ad hoc or human-driven workflows whose phase structure or near-term network supportability cannot be projected with sufficient confidence.

The proposed framework introduces a workflow-level coordination layer on top of existing QoS interaction and service exposure mechanisms, rather than replacing them. By combining bounded capability disclosure, workflow-level demand expression, and structured runtime adaptation, the framework enables forward-looking coordination across multiple future phases rather than isolated request-driven QoS handling.

\section{Coordination Framework}
\label{sec:framework}

\begin{figure*}[!t]
  \centering
  \includegraphics[width=0.9\textwidth]{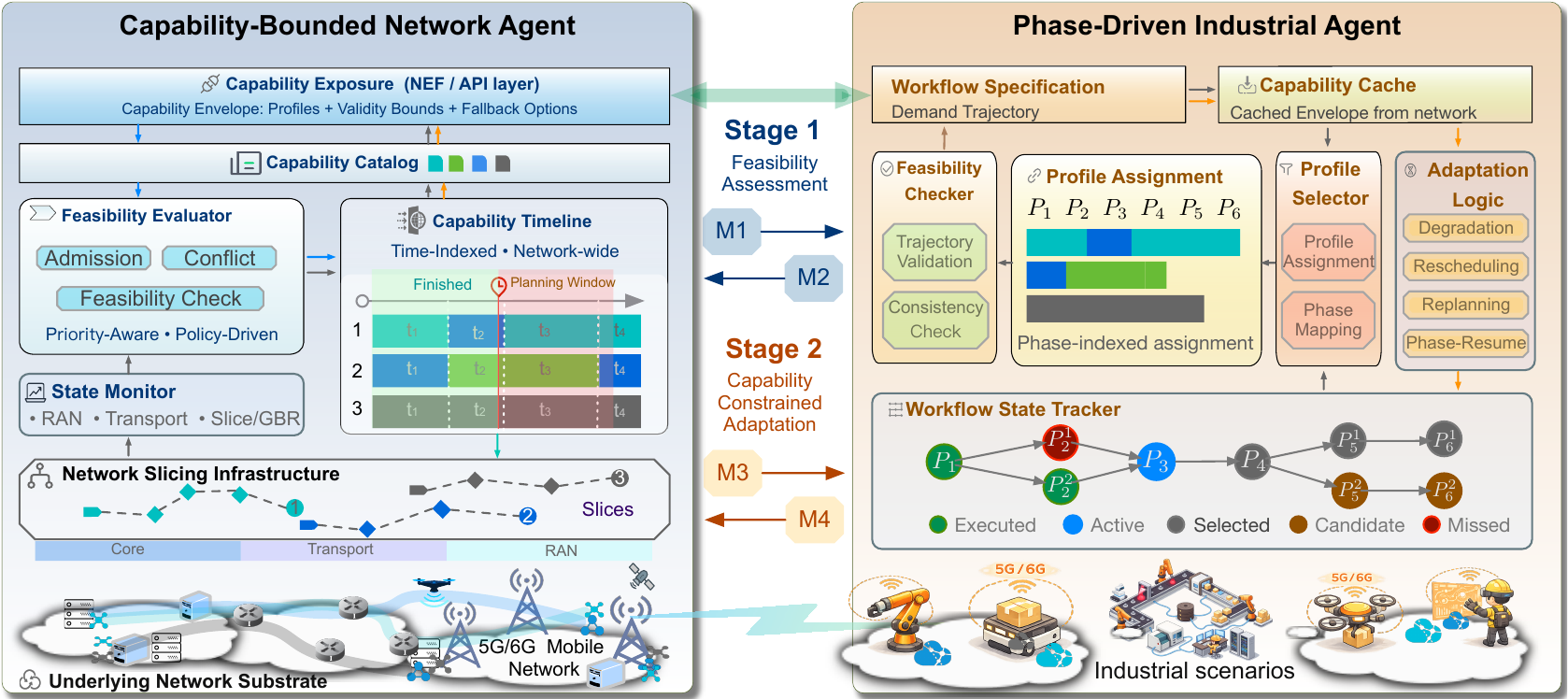}
  \caption{Capability-aware coordination framework with a capability-bounded network agent and a phase-driven industrial agent.}
  \label{fig:arch}
\end{figure*}

This section explains how the planning window, capability envelope, and demand trajectory are realized through the interaction between a network agent and industrial agents, and how a two-stage protocol maintains their alignment during workflow execution. Fig.~\ref{fig:arch} illustrates the architecture for one network agent and one industrial agent for clarity. In practice, one network agent can coordinate with multiple industrial agents, each managing one or more concurrent workflows.

\subsection{Network Agent: Capability Envelope Lifecycle}

The network agent is responsible for the generation, maintenance, and exposure of capability envelopes. Each capability envelope provides the industrial agent with a bounded view of the QoS profiles that can be sustained within the current planning window, without revealing internal resource allocation or scheduling details. In networking terms, the capability envelope is a scoped QoS supportability summary rather than a raw bandwidth value: it may reflect bandwidth, latency, jitter, reliability, validity bounds, and fallback options as supported by the underlying network domains. The capability envelope lifecycle comprises three functions: generation, consistency maintenance, and scoped exposure.

\emph{Capability Envelope generation.}  
The network agent continuously monitors network-side supportability inputs through the State Monitor, including RAN radio conditions and resource availability, slice and GBR budgets, transport or core processing margins, and policy constraints. Based on these inputs and committed demand trajectories, the Feasibility Evaluator projects which QoS profiles can be sustained over the planning window under current load. This projection forms an internal capability timeline, namely a time-bounded view of supportable profiles and their validity bounds across the planning window. Because this timeline already reflects admitted demand trajectories from multiple industrial agents, each newly derived capability envelope is consistent with existing commitments rather than being evaluated in isolation. Generating envelopes from a shared capability timeline therefore helps prevent over-commitment when multiple workflows share the same slice.

\emph{Capability Envelope consistency.}  
The capability timeline is updated whenever either side of the coordination problem changes. When a new demand trajectory is admitted, the corresponding commitment is recorded and the affected validity bounds are revised accordingly. When network conditions evolve, for example due to capacity reduction, resource release from neighboring workflows, or changes in radio quality, the Feasibility Evaluator re-assesses the active window. If a previously admitted trajectory segment can no longer be sustained, the network agent triggers Stage~2 coordination for the affected industrial agent. Consistency is thereby maintained proactively: the network agent detects emerging conflicts before they manifest as service disruption at subsequent phase transitions.

\emph{Capability Envelope exposure.}  
From the capability timeline, the network agent derives a capability catalog that organizes admissible QoS profiles and their projected validity intervals. The Capability Exposure function then scopes this catalog for each industrial agent, producing an agent-specific envelope that includes only the profiles and bounds relevant to that agent's planning window. Internal capacity planning, scheduling logic, cross-agent allocation decisions, and domain-specific resource states remain hidden. The exposed envelope thus provides the industrial agent with sufficient information to construct a feasible demand trajectory, while keeping network-internal decision making encapsulated.

\subsection{Industrial Agent: Demand Trajectory Lifecycle}

Each industrial agent manages one or more workflows and maintains a demand trajectory for each. The demand trajectory is a coordination object visible to the network agent: it encodes selected QoS profiles, phase ordering, expected durations, business priority, and permissions for adaptation. Its lifecycle comprises construction, validation, abstraction, and update.

\emph{Demand Trajectory construction.}  
Starting from the workflow phase structure maintained by the Workflow State Tracker and latest capability envelope cached after $M1$, the industrial agent assigns a QoS profile to each upcoming phase within the planning window. The Profile Selector chooses profiles that satisfy workflow requirements while remaining within the disclosed admissible set, taking into account degradation tolerance, business priority, and whether the workflow permits profile upgrades or downgrades at phase boundaries. The result is a draft profile assignment covering the full planning window.

\emph{Demand Trajectory validation.}  
The Feasibility Checker validates the draft assignment as a whole against the cached envelope, ensuring that the selected profile sequence remains consistent across the planning window. This workflow-level check avoids the failure mode of per-phase selection, in which profiles appear admissible individually but become infeasible once the full sequence is considered over overlapping time segments. Final confirmation remains with the network agent, but local validation helps avoid unnecessary renegotiation.

\emph{Demand Trajectory abstraction.}  
Before submission, the industrial agent abstracts the validated assignment into a demand trajectory suitable for network-side evaluation. The trajectory retains phase ordering, expected durations, selected profiles, business priority, and upgrade or degradation permissions. It hides workflow-internal structure, including alternative execution paths, local decision rationale, and proprietary process logic. This abstraction ensures that the network agent receives the information needed for feasibility assessment and resource configuration, but no more. For example, the inspection trajectory contains idle, routine patrol, active inspection, and cooldown segments, each with an expected duration, preferred profile, lower admissible profile, and adaptation permission. The capability envelope then indicates which profiles remain supportable within the corresponding validity intervals.

\emph{Demand Trajectory update.}  
When Stage~2 is triggered by a capability notification ($M3$), the industrial agent's Adaptation Logic evaluates the impact of the change on the phases of the demand trajectory. Depending on workflow policy, it may accept a lower admissible profile, temporarily downgrade a non-critical phase to protect a higher-priority one, defer an affected phase to a later interval where capacity is expected to recover, or replan the remaining phase sequence altogether. If resources become temporarily insufficient, the framework therefore supports controlled fallback rather than abrupt service failure. When capacity is restored, previously degraded segments may also be resumed under updated constraints. The revised demand trajectory is then re-validated against the updated capability envelope and resubmitted through $M4$.

\subsection{Two-Stage Coordination}

\begin{figure}[!t]
  \centering
  \includegraphics[width=0.4\textwidth]{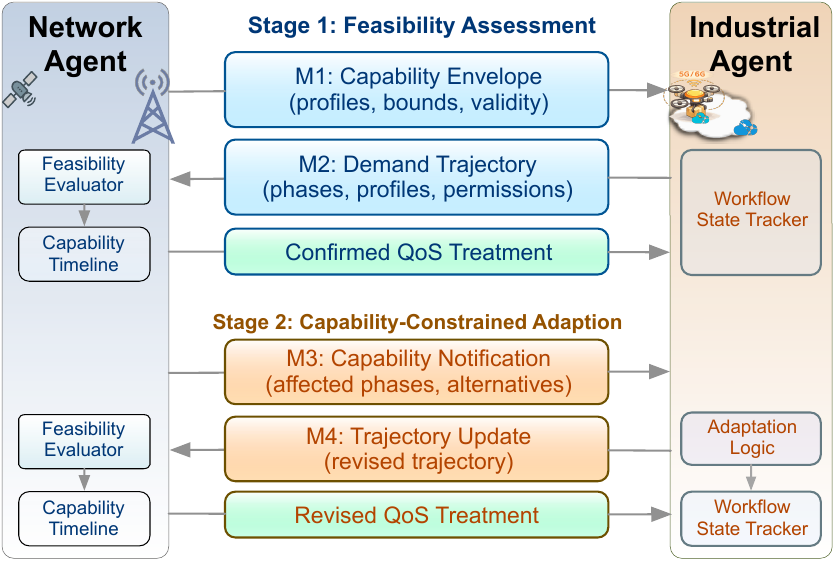}
  \caption{Two-stage coordination workflow between the network agent and industrial agent.}
  \label{fig:workflow}
\end{figure}

The coordination protocol operates within the planning window, a rolling window whose length is set by the network agent according to how far ahead network capability can be projected with sufficient confidence. The window typically spans multiple workflow phases and is long enough to cover at least the next critical phase transition. As time advances, expired segments are removed and newly projected capability is appended. Industrial agents update their demand trajectories accordingly to maintain alignment within the active window. This rolling process provides continuous coordination without requiring explicit session restart or repeated full renegotiation. Accurate capability prediction enables precise planning and demand-trajectory management within the planning window, while inaccurate prediction may cause suboptimal profile assignment and more frequent runtime updates, such as downgrades, upgrades, or trajectory revisions. Therefore, the planning-window length, capability-update threshold, and notification policy should balance coordination accuracy, prediction stability, and signaling overhead; $M3$ is triggered mainly when validity bounds are crossed or admitted trajectories are affected. Fig.~\ref{fig:workflow} illustrates the two stages operating within this window.

\subsubsection{Stage 1: Feasibility Assessment}

Stage~1 establishes initial alignment between projected network capability and workflow demand. It begins with message $M1$, in which the network agent discloses the capability envelope for the current planning window. The envelope specifies which QoS profiles the network can sustain, together with their validity periods and associated capacity bounds. These bounds reflect projected capability under current resource allocation, existing service commitments, and expected near-term capacity evolution.

Upon receiving $M1$, the industrial agent caches the capability envelope and constructs a demand trajectory following the lifecycle described above. It assigns a profile to each upcoming phase, validates the sequence against the cached envelope, and abstracts the result for submission. The demand trajectory is sent in $M2$, carrying phase ordering, expected durations, selected QoS profiles, business priority, and permitted adaptation ranges. This allows the network agent to assess whether the demand sequence can be sustained over the planning window, rather than evaluating each phase independently.

The network agent evaluates $M2$ against available resources and existing commitments within the same planning window. If the proposed demand trajectory is feasible, the network agent confirms the selection, records the demand trajectory in the capability timeline, and applies the corresponding QoS treatment through slicing and policy control mechanisms. The industrial agent can then proceed with workflow execution under a jointly assessed plan. If the demand trajectory cannot be fully supported, the network agent identifies the affected phases or time segments and moves to Stage~2.

\subsubsection{Stage 2: Capability-Constrained Adaptation}

Stage~2 restores alignment when network capability changes after initial confirmation, or when a previously confirmed demand trajectory can no longer be sustained. This may happen because capacity has dropped, a higher-priority workflow has been admitted, or radio conditions have deteriorated.

When such a change is detected, the network agent sends $M3$, a capability notification that identifies the affected phases or time segments for which the current profile assignment is no longer supportable. The notification may also include admissible alternatives that remain feasible under updated conditions. These alternatives do not prescribe a single response; instead, they provide the industrial agent with a bounded capability view for coordinated adaptation. In some cases, no admissible alternative is available for the affected phase or time segment.

Upon receiving $M3$, the industrial agent evaluates the impact on workflow execution using its Adaptation Logic. Depending on workflow policy and the adaptation range declared in the original demand trajectory, it may accept a lower profile, downgrade a less critical phase to preserve a higher-priority one, defer a delay-tolerant activity to a later slot, or replan the remaining workflow altogether. The revised demand trajectory is re-validated against the updated capability envelope and returned in $M4$.

Upon receiving $M4$, the network agent verifies that the updated demand trajectory is feasible under current conditions, applies the revised QoS configuration through slicing and policy control mechanisms, and updates the capability timeline accordingly. When a capacity reduction affects multiple industrial agents, the network agent determines the order of adaptation based on the business priority declared in each demand trajectory. If the same capability change affects other industrial agents, separate $M3$ notifications are issued to each, enabling concurrent adaptation without exposing cross-agent internals. In this way, temporary resource shortfalls are handled through coordinated fallback and later recovery, rather than abrupt rejection at the next phase transition.

From an implementation perspective, $M1$--$M4$ can be exposed through NEF or a similar API layer, while PCF, NWDAF, and Alternative QoS Profiles may respectively support policy enforcement, prediction inputs, and fallback candidates. The coordination layer organizes these functions around workflow-level capability and demand objects.

\subsection{Summary}

The key difference between coordinated and request-driven operation is not whether QoS profiles can be enforced, since both approaches may rely on the same slicing and policy control mechanisms. The difference lies in when and how feasibility is assessed. Under request-driven operation, each phase transition is evaluated locally and independently. Under the proposed framework, feasibility is coordinated across a bounded sequence of upcoming phases within the planning window. Stage~1 establishes this workflow-level alignment, while Stage~2 maintains it as network conditions evolve. The result is a shared temporal scope within which both sides can plan, adapt, and recover.

\section{Case Study}
\label{sec:experiment}

We illustrate and evaluate the framework through two complementary studies. First, a deployment on an operational 5G/B5G testbed demonstrates how the two-stage coordination protocol operates in practice, including planning-stage alignment and runtime adaptation under both anticipated and unanticipated capacity changes. Second, a large-scale simulation examines how the same coordination logic behaves when many concurrent workflows compete for shared network capacity.

The baseline is per-phase request-driven QoS interaction, not an enhanced reactive design. Predictive QoS notification and Alternative QoS Profiles serve as session- or fallback-scoped mechanisms; the gains here reflect bounded workflow-level coordination on top of existing QoS enforcement.

\subsection{5G Testbed Deployment}
\label{sec:testbed}

\begin{figure}[!t]
\centering
\includegraphics[width=3.4in]{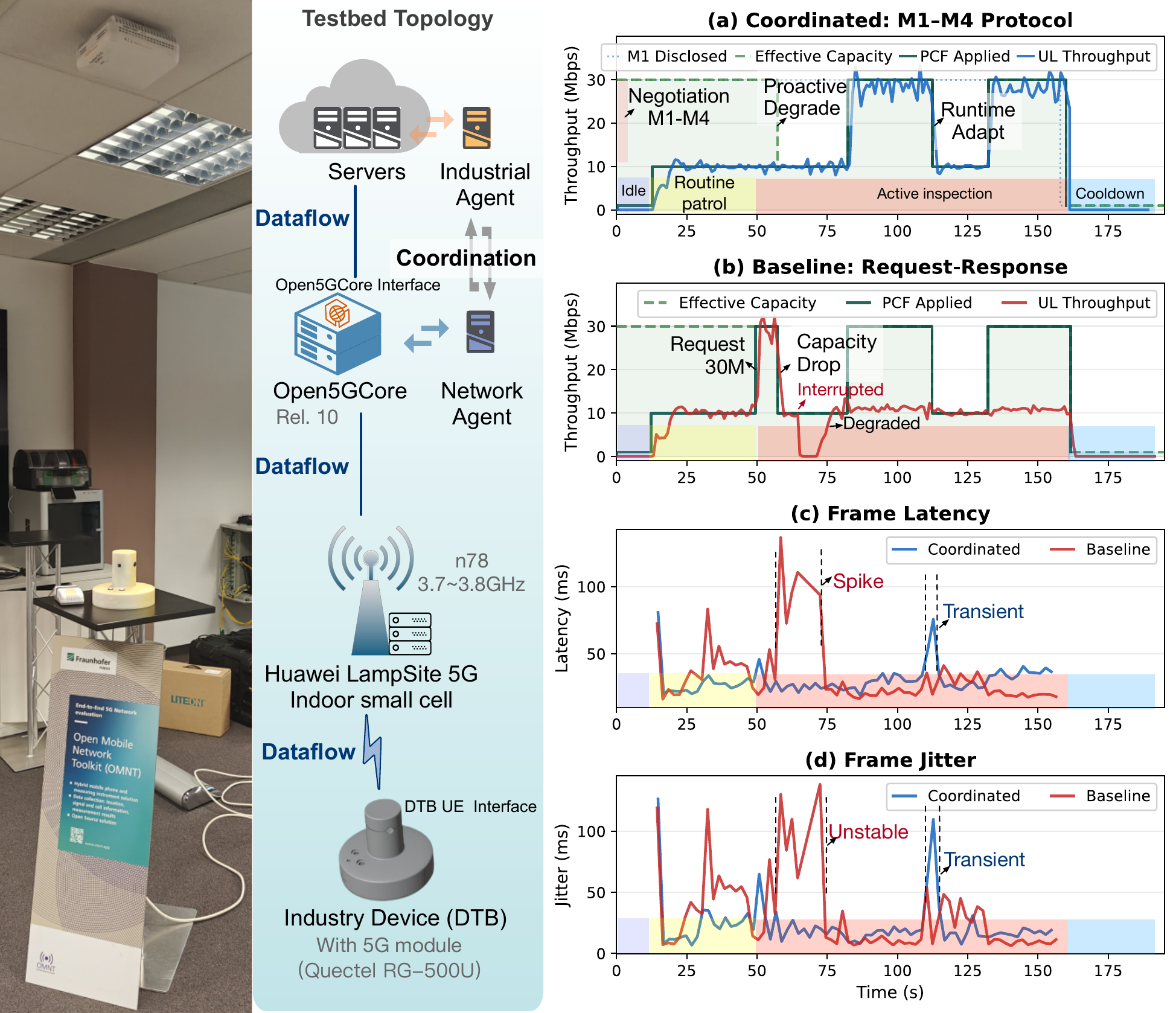}
\caption{5G testbed result. Left: testbed topology. Right: coordinated and request-driven operation showing (a)--(b) throughput with capacity, (c) frame latency, and (d) frame jitter.}
\label{fig:testbed_timeline}
\end{figure}

The first study involves a video-based industrial inspection workflow running on a 5G/B5G testbed. The workflow is executed by a Digital Twin Box (DTB)~\cite{iotm_dtb} equipped with cameras and proceeds through four phases: idle (1~Mbps, 10\,s), routine patrol (10~Mbps, 37\,s), active inspection (30~Mbps, 110\,s), and cooldown (1~Mbps, 30\,s). This workflow is representative of industrial services whose communication requirements change markedly across phases, making it well suited for comparing coordinated and request-driven operation.

The network side is implemented on the 6G-SANDBOX Berlin platform~\cite{6gsandboxBerlinPlatform} hosted at the Fraunhofer FOKUS 5G infrastructure, integrating a Huawei LampSite 5G indoor small cell operating in the n78 band at 3.7--3.8~GHz with 100~MHz uplink/downlink bandwidth, 30~kHz subcarrier spacing, and a 4T4R configuration, together with the Open5GCore core network. Three Guaranteed Bit Rate (GBR) QoS profiles (1, 10, and 30~Mbps, all with 5G QoS Identifier (5QI)\,=\,4) are pre-configured and enforced through the PCF. The 5QI identifies the QoS treatment associated with these GBR profiles. In this testbed, the capability envelope is instantiated through PCF-enforced GBR service budgets, while the radio link is kept stable and is not used as the varying bottleneck. Coordination messages ($M1$--$M4$) are exchanged over Message Queuing Telemetry Transport (MQTT), while profile changes are applied through the PCF. These implementation details provide a concrete realization of the framework, but the main purpose of the case study is to show how coordination changes workflow behavior under varying network supportability.

\emph{Scenario design.}
The nominal service budget for the inspection workflow is core-network enforced and configured as 30~Mbps. During 55--80\,s, admitted background traffic occupies 20~Mbps of this budget, reducing the effective available service capacity to 10~Mbps during part of the active inspection phase. This creates a foreseeable mismatch between workflow demand and network capability. In addition, an unanticipated service-budget reduction to 10~Mbps is injected at 110\,s, followed by recovery at 130\,s. The coordinated and request-driven modes are exposed to the same supportability conditions; only the coordination mechanism differs.

\emph{Coordinated operation.}
In coordinated mode, the industrial agent first submits a demand trajectory for the inspection workflow, and the network agent performs Stage~1 feasibility assessment against the disclosed capability envelope. For the considered scenario, this assessment shows that the 30~Mbps profile cannot be sustained throughout the active inspection phase because of the known contention interval. The interaction therefore proceeds immediately to Stage~2 capability-constrained adaptation, in which the network agent indicates a lower admissible profile for the affected segment and the industrial agent updates its demand trajectory accordingly. As a result, when contention begins at $t=55$\,s, the DTB is already operating within the supportable limit, and the foreseeable capacity reduction causes no visible stream disruption.

Stage~2 is invoked again during execution whenever network supportability changes. When the contention ends at $t=80$\,s, the network agent notifies the industrial agent that the higher profile is feasible again, and the DTB resumes full-rate streaming within 2.2\,s. When the unanticipated drop occurs at $t=110$\,s, the same capability-constrained adaptation mechanism triggers bounded fallback to the lower profile, again within 2.2\,s. Once capacity recovers at $t=130$\,s, the DTB returns to the higher profile. As shown in Fig.~\ref{fig:testbed_timeline}(a), these transitions remain smooth and the video stream continues without interruption. The end-to-end adaptation delay includes approximately 2.0\,s of PCF API latency for applying the updated QoS treatment.

\emph{Request-driven operation.}
Under request-driven operation, the industrial agent requests 30~Mbps when active inspection begins and is initially accepted. However, when available capacity later drops to 10~Mbps, the DTB continues transmitting above the enforceable limit because no capability notification mechanism exists. Packet loss accumulates and the stream interrupts at $t=66$\,s. The industrial agent then retries 30~Mbps, receives rejection, and eventually falls back to 10~Mbps, with service resuming only after an 8.2\,s interruption. Because the baseline has no structured notification when capacity later improves, two subsequent 30~Mbps windows remain unused even though higher-rate operation becomes feasible again.

\emph{Observed benefit.}
Table~\ref{tab:testbed_results} summarizes the outcomes. Coordinated operation improves mean throughput during active phases by 55\% by exploiting both high-capacity windows through explicit capability notification and demand trajectory revision. Frame-level analysis in Fig.~\ref{fig:testbed_timeline}(c)--(d) shows lower latency and jitter. More importantly, the case study highlights the operational value of the framework: Stage~1 prevents foreseeable mismatch before it affects execution, while Stage~2 restores alignment when network conditions change unexpectedly. The resulting gain is not only higher throughput, but also improved service continuity throughout the workflow.

\begin{table}[!t]
\centering
\caption{Testbed comparison: coordinated versus request-driven operation.}
\label{tab:testbed_results}
\begin{tabular}{p{0.44\linewidth} p{0.22\linewidth} p{0.22\linewidth}}
\hline
\textbf{Metric} & \textbf{Coordinated} & \textbf{Request-driven} \\
\hline
Stream interruptions & 0 & 1 (8.2\,s) \\
Mean throughput (active) & 16.4\,Mbps & 10.5\,Mbps \\
P95 frame latency & 67.3\,ms & 111.0\,ms \\
Mean frame jitter & 22.6\,ms & 31.2\,ms \\
Stage~2 adaptation time & $\sim$2.2\,s & N/A \\
\hline
\end{tabular}
\end{table}

\subsection{Multi-Workflow Simulation}
\label{sec:simresults}

\begin{figure}[!t]
\centering
\includegraphics[width=3.4in]{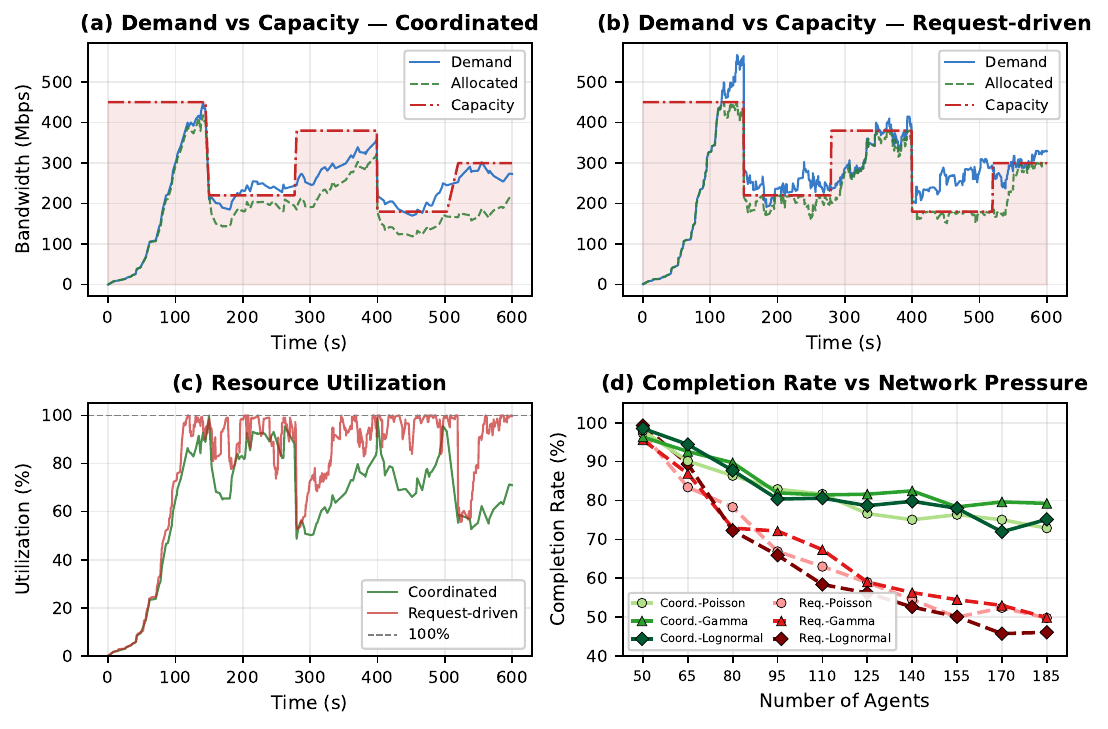}
\caption{Simulation results. (a)--(b) Aggregate demand and capacity under coordinated and request-driven operation (120 agents). (c) Utilization comparison. (d) Completion rate under Poisson, Gamma, and Lognormal arrival models.}
\label{fig:simulation}
\end{figure}

The testbed case study illustrates the coordination workflow for a single industrial service. We now examine the same framework under a shared-network setting with many concurrent workflows competing for limited service capacity. The objective is not to reproduce every aspect of the testbed, but to demonstrate how the proposed coordination logic behaves as network pressure increases and per-request interaction becomes increasingly unreliable.

The simulation models up to 185 industrial agents, each executing workflows composed of phases whose requirements are drawn from a five-level QoS set (1, 5, 10, 20, and 30~Mbps). Three workflow classes are considered: critical inspection (10\%), routine monitoring (60\%), and background sensing (30\%). Fig.~\ref{fig:simulation}(a)--(c) and Table~\ref{tab:sim_standard} report the Poisson-arrival reference case. Fig.~\ref{fig:simulation}(d) further compares Poisson, Gamma, and Lognormal arrival models under the same average offered load. Network capacity evolves according to a heavy traffic scenario (450$\rightarrow$220$\rightarrow$380$\rightarrow$180$\rightarrow$300~Mbps), capturing fluctuations caused by concurrent service demand. As in the testbed study, the comparison is between coordinated operation and a request-driven baseline that evaluates QoS independently at each phase transition, without workflow-level feasibility assessment, capability envelope disclosure, or workflow-level adaptation.

\emph{Effect of coordination under shared load.}
Under these conditions, the key difference is how infeasibility is handled. In the request-driven baseline, infeasible requests reach the network at phase boundaries and are rejected there. In the coordinated framework, infeasibility is addressed earlier through demand trajectory revision inside the planning window. This difference changes the system behavior: many workflows that would otherwise fail completely can instead continue in degraded form and still reach completion.

This effect is reflected in Table~\ref{tab:sim_standard}. At 120 agents, the coordinated framework achieves a 71.5\% workflow completion rate, compared with 45.0\% for the request-driven baseline. The coordinated framework eliminates hard rejections, converting would-be failures into degraded but completed workflows through capability notification and controlled fallback.

\begin{table}[!t]
\centering
\caption{Simulation results at 120 agents: coordinated versus request-driven operation.}
\label{tab:sim_standard}
\begin{tabular}{p{0.44\linewidth} p{0.22\linewidth} p{0.22\linewidth}}
\hline
\textbf{Metric} & \textbf{Coordinated} & \textbf{Request-driven} \\
\hline
Total workflows & 389 & 389 \\
Optimal completion & 203 & 137 \\
Degraded completion & 75 & 38 \\
Failed & 111 & 214 \\
Completion rate & 71.5\% & 45.0\% \\
Network-side hard rejections & 0 & 214 \\
\hline
\end{tabular}
\end{table}

\emph{Aggregate demand and utilization.}
Fig.~\ref{fig:simulation}(a)--(b) illustrates the aggregate relationship between workflow demand and available capacity. With coordination, workflow demand is revised toward feasible profile combinations inside the planning window, so the aggregate demand remains better aligned with the capacity. In the baseline, phase transitions repeatedly generate requests that exceed instantaneous supportability, producing sharp mismatch and widespread rejection.

Fig.~\ref{fig:simulation}(c) compares effective utilization. The coordinated framework utilizes available capacity more effectively because it redirects workflows toward feasible demand trajectories instead of discarding infeasible requests. This is particularly valuable in stressed intervals, where degraded operation preserves partial service capability and keeps capacity productive.

\emph{Effect of increasing load.}
Fig.~\ref{fig:simulation}(d) presents the effect of increasing load from 50 to 185 agents. As load increases, both approaches degrade, but they do so differently. The request-driven baseline drops rapidly once hard rejections become frequent. By contrast, the coordinated framework declines more gradually because it maintains workflow completion through bounded fallback and trajectory revision. The gap therefore widens as shared-network pressure increases. Across the Poisson, Gamma, and Lognormal pressure sweeps, the coordinated framework maintains the same qualitative advantage; at 185 agents, the completion-rate gains are 23.1, 29.4, and 29.0 percentage points, respectively.

The testbed and simulation are complementary rather than directly comparable. The testbed demonstrates $M1$--$M4$ under controlled PCF-enforced service-budget changes, while the simulation abstracts the same coordination principle under many concurrent workflows. In both settings, the gain comes from adapting before infeasible phase requests become hard failures.

\section{Conclusion and Future Work}
\label{sec:conclusion}

Industrial services in 5G and 6G networks increasingly operate as phase-evolving workflows, while network capability also changes over time. This article presented a capability-aware coordination framework based on a planning window, a capability envelope, and a demand trajectory, enabling forward-looking feasibility assessment and runtime adaptation across upcoming phases.

The 5G testbed and multi-workflow simulation provide complementary evidence that coordinated operation improves service continuity and workflow completion over request-driven operation. The results support capability-aware QoS coordination beyond reactive per-request handling.

Future work will extend the framework toward 6G-oriented deployment, including AI-assisted capability prediction, mobility-driven handover, dynamically composed workflows, and multi-domain coordination. Practical deployment will also require careful tuning of capability-envelope granularity and update frequency to balance coordination precision and signaling overhead.

\section{Acknowledgments}
\noindent {This work was supported in part by the Federal Ministry of Research, Technology and Space (BMFTR), Germany, through the Project 6GEM+ under Grant 16KIS2411; and by the European Union's Horizon Europe programme through 6G-Path (Grant No.~101139172) and 6GSandbox (Grant No.~101096328). This research was also partially conducted at ICTFICIAL Oy. The paper reflects only the authors’ views.}

\bibliography{bibtex/IEEEabrv,bibtex/mag}

\vfill

\end{document}